\documentclass[preprint]{aastex}
\usepackage{natbib}
\shorttitle{Hot Disk around BD+20 307}
\shortauthors{Weinberger et al.}

\newcommand{\lir}{L$\rm _{IR} / L_\star$}

\begin{document}

\title{The Absence of Cold Dust and the Mineralogy and Origin of the Warm Dust Encircling BD +20 307}

\author{A. J. Weinberger}
\affil{Department of Terrestrial Magnetism, Carnegie Institution of
Washington}
\affil{5241 Broad Branch Road NW, Washington, DC 20015}
\email{weinberger@dtm.ciw.edu}

\author{E. E. Becklin}
\affil{Department of Physics and Astronomy, University of California Los Angeles}
\affil{Box 951547, Los Angeles, CA 90095-1562}
\email{becklin@astro.ucla.edu}

\author{I. Song}
\affil{Department of Physics and Astronomy, The University of Georgia}
\affil{Athens, GA 30602-2451}
\email{song@uga.edu}

\author{B. Zuckerman}
\affil{Department of Physics and Astronomy, University of California Los Angeles}
\affil{Box 951547, Los Angeles, CA 90095-1562}
\email{ben@astro.ucla.edu}

\begin{abstract}

Spitzer Space Telescope photometry and spectroscopy of BD +20 307 show that
all of the dust around this remarkable Gyr-old spectroscopic binary arises
within 1 AU. No additional cold dust is needed to fit the infrared
excess. Peaks in the 10 and 20 $\mu$m spectrum are well fit with small
silicates that should be removed on a timescale of years from the system.
This is the dustiest star known for its age, which is $\gtrsim$1 Gyr.
The dust cannot arise from a steady-state collisional cascade.  A catastrophic collision of two rocky,
planetary-scale bodies in the terrestrial zone is the most likely source for this
warm dust because it does not require a reservoir of planetesimals in the
outer system.
 
\end{abstract}

\section{Introduction}

A number of extreme collisions may have shaped the final states of the
terrestrial planets.  Impact by a Mars-sized progenitor on the early Earth,
perhaps 30 Myr after the Solar System formed, could have created the Moon
\citep[e.g.][]{Canup:2004}. Other collisions may have stripped Mercury's mantle
\citep{Benz:1988} or caused the hemispheric crustal thickness asymmetry of Mars
\citep{Marinova:2008}.  Subsequent bombardment of the planets may not have
generated planetary scale changes, but the ``Late Heavy
Bombardment,'' an era approximately 600 Myr after Solar System formation,
either culminated a long sequence of bombardments or resulted in a burst of
impacts on the terrestrial planets \citep{Gomes:2005,Strom:2005}.

Evidence of cataclysmic impacts in other planetary systems is rare. Enhanced
collisions should result in the production of copious small particles,
i.e. dust, which can be observed by their re-radiation of stellar light
\citep{Kenyon:2004}.  Stars older than 100 Myr with substantial amounts of
warm dust are rare, and at most a few percent of field stars have any
detectable warm dust at all
\citep[e.g.][]{Meyer:2008,Trilling:2008,Greaves:2009,Koerner:2010}.  At any
given age, stars that are dustier than average could be interpreted as having
had a recent, stochastic, event, but many could also be the result of
differing initial masses and configurations of planetesimals.  Only the
extremely dusty disks at any given age must be from giant collisions.

BD+20 307 is a relatively anonymous field star at 96$^{+13}_{-10}$ pc
\citep{Leeuwen:2007} from the Sun. It is actually a tidally locked
spectroscopic binary with a period of 3.5~d composed of two nearly identical
late F-type stars with a mass ratio of 0.91 \citep{Weinberger:2008}.
\citet{Weinberger:2008} and \citet{Zuckerman:2008} investigated the stellar
properties, in particular the weak and disparate Lithium abundances in the two
stars, and estimated an age $\gtrsim$ 1 Gyr. At that age, its dust opacity,
estimated by L$_{\rm IR}$/L$_*$ to be 0.033 \citep{Song:2005}, is orders of magnitude larger than
every other star of its age. In fact, such dustiness would be extraordinary
even for a star 10--100 times younger.

A small class of very (warm) dusty objects now exists in the literature (see
summaries in \citet{Rhee:2008,Moor:2009}).  The extreme examples of these
 and also some disks with strange infrared spectra such as HD
172555 \citep{Lisse:2009} and 51 Oph; \citep{Ancker:2001}, have been
attributed to impacts of planet-sized bodies.

Here, we investigate the unusually dusty system around the star BD+20
307 using the Spitzer Space Telescope. These observations constrain the
size distribution and location of a very large amount of dust and
provide perhaps the best evidence yet for large-scale collisions in
another system.

\section{Observations and Data Reduction}

We observed BD +20 307 with Spitzer with all three instruments -- IRAC, MIPS,
and IRS. An observing log is given in Table \ref{table_obslog}.

\begin{deluxetable}{lclcc}
\tablecaption{Summary of Observations and Photometry\label{table_obslog}}
\tablewidth{0pt}
\tablehead{
\colhead{Instrument} 
        &\colhead{Band}
                       &\colhead{Date}
                                    &\colhead{Integr. time (s)}
                                          &\colhead{Photometry (mJy)}}
\startdata
IRAC    &1 (3.5$\mu$m) &2005-08-20  &0.64  & 285.0 $\pm$ 7.3 \\
IRAC    &2 (4.5$\mu$m) &2005-08-20  &0.64  & 210.4 $\pm$ 4.8 \\
IRAC    &3 (5.7$\mu$m) &2005-08-20  &0.64  & 165.3 $\pm$ 12.0 \\
IRAC    &4 (7.9$\mu$m) &2005-08-20  &0.64  & 410.2 $\pm$ 22.5\tablenotemark{a}\\
MIPS    &1 (24$\mu$m)  &2007-01-21  &2.62  & 441.0 $\pm$ 0.8\\
MIPS    &2 (70$\mu$m)  &2007-01-21  &167.8 & 28.6 $\pm$ 1.9\\
MIPS    &3 (160$\mu$m) &2007-01-21  &209.8 & $<$22\\
IRS &0 (5.2--14.5$\mu$m)&2006-01-15 &12.6  &--\\ 
IRS &1 (9.9--19.6$\mu$m)&2006-01-15 &188.7 &--\\
IRS &3 (18.7--37.2$\mu$m)&2006-01-15&1219  &--\\
\enddata
\tablenotetext{a}{Not color corrected}
\end{deluxetable}

\subsection{IRAC}

IRAC data were taken in subarray mode, which yields 64 individual images. The
S14 pipeline reduction was used, and the ``Basic Calibrated Data'' (BCD)
images were mosaiced with the Spitzer Science Center's MOPEX software. There
are two photometric corrections to consider. The first is the ``Pixel phase
dependent photometric correction'' that corrects the pixel response function
for where the peak response is within a pixel. This correction was applied to
channel 1, and amounted to 1.7\%.  The second is the array location dependent
photometric correction, due to the fact that IRAC is flat-fielded on the
Zodiacal spectrum. From the overall shape of the BD +20 307 spectral energy
distribution (SED), we know that its spectrum is shallower in slope than Rayleigh-Jeans
from 3 --7 $\mu$m
but not as flat as the Zodiacal emission. The correction was applied at
Channel 1, and amounted to 0.7\%. It was also applied to channels 2 and
3, but made less of an effect, and was not applied for channel 4.
Aperture photometry was performed in the radius 2, 3, and 5 pixel radii
apertures given in the IRAC handbook, and the standard deviation of these
taken as the measurement uncertainty.

Finally, the IRAC data need to be color corrected from the default $\nu F_\nu
= constant$ calibration to the true BD+20 307 spectrum. This effect is small
for Bands 1 and 2, where the source spectrum is approximately a 1000 K
blackbody, and the correction is 0.7\%.  For bands 3 and 4, we integrated the
observed IRS spectrum over the IRAC filter bandpasses. This yields a
correction of 3\% at Band 3 and 71\% for Band 4. The Band 4 correction is
unreasonably large; the difference between the integral under the IRS spectrum
and the IRAC Band 4 (un-color corrected) flux density is only 26\%. When
integrated, the very large spectral slope at 8 $\mu$m due to the silicate
feature that peaks at $\approx$10 $\mu$m, causes a large change in color
correction for any small change in spectral transmission response of the
system (telescope + filter, etc.). This transmission does not appear to be
well enough known to provide an accurate color correction; therefore, the Band
4 flux was not used for further analysis.

\subsection{MIPS}

At 24 $\mu$m, the pipeline produced post-BCD mosaic was used for aperture
photometry using a 6 pixel aperture and corrections given in \citet{Su:2006}.
At 70 $\mu$m, MOPEX was used to mosaic the pipeline filtered BCDs. Aperture
photometry was performed with the 35'' aperture and corrections from
\citet{Gordon:2007}.  At 160 $\mu$m, the pipeline mosaic was used. Because at
this band no source is seen at the nominal location or elsewhere in the field,
an upper limit to the source flux was placed by doing a raster of aperture
positions on the image around the nominal location and calculating three times
the standard deviation of these.

\subsection{IRS}

For all channels, the pipeline version 14 data were used. The short-low (5.2
-- 14.5 $\mu$m) BCD data were cleaned with IRS\_CLEAN software to remove bad
pixels. Each cleaned BCD was sky subtracted with uncertainties propagated in
quadrature. The spectra were extracted with the Spitzer Science Center's SPICE
software using the standard extraction parameters.

The short-high (9.9 -- 19.6 $\mu$m) BCD data were cleaned with IRS\_CLEAN and
the unaltered campaign pixel mask. The resulting cleaned BCDs were medianed to
make a final spectrum that was extracted with SPICE and the standard
extraction parameters.  The two nods were averaged together. Orders were
trimmed to eliminate as many points with undefined uncertainties as possible.

For the long-high (18.7 -- 37.2 $\mu$m) data, separate observations on nearby sky were taken
to help in elimination of rogue pixels. A median sky frame was created
with the standard deviation of the 20 sky files as its uncertainty. This
frame was then subtracted from each of the object spectra, with
uncertainties propagated.

The campaign rogue mask was augmented by any pixels that were flagged in 15/20
of the individual spectra. The combined mask was then used to clean the sky
subtracted object spectra. The final cleaned BCDs for each nod were median
combined and then extracted with SPICE. The spectra from the two nods, which
agree to within their statistical uncertainties, were averaged.

The spectra from all channels were combined by using their overlap
regions -- short-high to short-low spectrum in the region
11.5--14 \micron\ and long-high to short-high at 19.02--19.47 \micron.
The complete spectrum was then normalized to the MIPS channel 1
photometry by integrating the spectrum under its transmission curve. A
check that this process works well is that integrating the final
normalized spectrum  under the IRAC channel 3 filter yields 155 mJy
compared with the measured 165 $\pm$ 12 mJy from the photometry reported in
Table \ref{table_obslog}.

\section{Results}

\begin{figure}
\center \includegraphics[angle=90,scale=0.7]{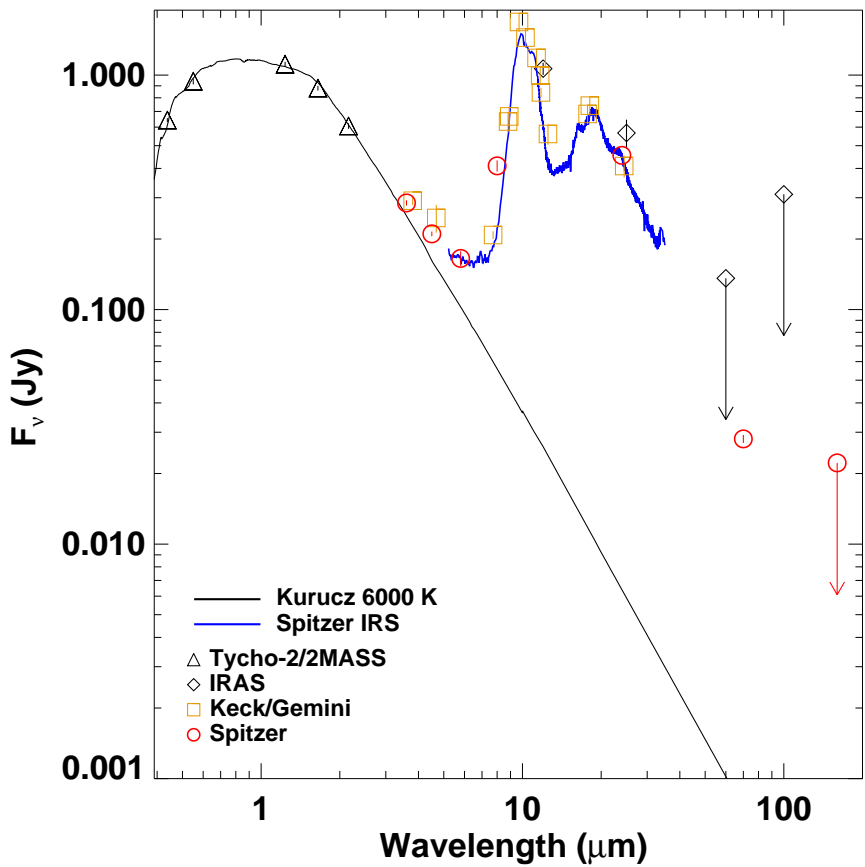}
\figcaption{Spectral Energy Distribution of BD+20 307. The Tycho-2 and 2MASS
  points (black triangles) were used to fit the Kurucz photosphere (black
  line). All long wavelength data are plotted including our Spitzer photometry
  (red circles) and spectroscopy (blue line) as well as IRAS (black diamonds)
  and older ground-based data from Song et al. 2005 (gold
  squares). \label{fig_sed}}
\end{figure}

Results of the new Spitzer photometry are given in Table \ref{table_obslog}.
As first reported in \citet{Song:2005}, the spectral energy distribution of
BD+20 307 at wavelengths longer than 5 \micron\ is dominated by emission by
small silicate particles. We show the overall SED including the new Spitzer
spectrum and photometry along with previous ground-based photometry in Figure
\ref{fig_sed}.

\subsection{Photosphere Modeling}

\begin{deluxetable}{lrrr}
\tablewidth{0pt}

  \tablecaption{Photosphere subtracted flux densities and model fits in mJy\label{table_photometry}}
  \tablehead{
\colhead{Band}
&\colhead{Photosphere}
       &\colhead{Disk flux}
                       &\colhead{Model (mJy)}}
  \startdata
  IRAC1    &261.0 &24.0  & 9.3\\
  IRAC2    &167.0 &43.4  &28.8\\
  IRAC3    &107.1 &58.2  &70.8\\
  IRAC4    & 60.7 &349.5 &293.2\\
  MIPS1    &  6.8 &434.2 &424.4\\
  MIPS2    &  0.8 & 27.8 & 40.2\\
  MIPS3    &  0.3 &$<$22 & 8.7 \\
  \enddata
\end{deluxetable}

We model the combined flux density of the nearly identical central stars with a single
Kurucz model atmosphere fit to the Tycho-2 and 2MASS photometry at 0.4 -- 2.2
\micron. The best fit model has a gravity of log(g)=4.5, T$_{\rm eff}$=6000,
and, at the parallactic distance of 96 pc, a luminosity of 1.94
L$_\odot$. \citet{Zuckerman:2008} finds the luminosity ratio of the two stars
as 0.78, which would imply individual luminosities of 1.10 and 0.84 L$_\odot$.
Table \ref{table_photometry} gives the amount of photosphere in each Spitzer
band and the remaining disk flux. The model photosphere was also subtracted
from the IRS spectrum.

\subsection{Spectral Fitting \label{section_fitting}}

We proceed by assuming a simple model for the composition and
location of the circumstellar dust and asking whether it is a plausible match
to the data. The presence of both broad and sharp spectral features indicates
emission by both amorphous and crystalline materials \citep[e.g.][]{Campins1989}.

The shortest observed peak, at 9.7 $\mu$m, is matched well by the wavelength
of the peaks in measurements of glassy silicates, although its width is
narrower than that produced by even very small ($\lesssim$ 0.1 $\mu$m) grains.
The breadth of the 16-20 $\mu$m peak is better fit by larger, 1 $\mu$m grains.
We chose to set the minimum grain size to 0.5 $\mu$m, which is approximately
the blow-out size for radiation pressure from the stars.  We then use a power
law distribution of sizes rising from 0.5 $\mu$m with slope -3.5, but the fit
does not materially change whether the power law distribution or just the
minimum-sized grains are used.  Smaller grains result in narrower peaks
that improve the fit at 10 $\mu$m but make it worse at 20 $\mu$m.

We take the simple model to be dust at a single distance from the star and
composed of four types of particulates: small amorphous dust of olivine
composition (MgFeSiO$_4$; hereafter called amorphous olivine), small
crystalline olivine (forsterite; Mg$_2$SiO$_4$)), small amorphous pyroxenes
(Mg$_{0.5}$Fe$_{0.5}$SiO$_3$), and large grains (i.e. those that behave as
blackbodies). These are common constituents used to model solar system comets
and circumstellar disks \citep[e.g][]{Li:2004}. 

All small particles (size $<<$ wavelength of observation) have absorption
efficiencies that fall faster than Rayleigh-Jeans (i.e. $\propto
\lambda^{-2}$) at wavelengths much larger than the particle size. Therefore,
some bigger grains are necessary to match the measured flux densities at
20--70 $\mu$m.  Again, as a matter of taste, we have chosen to keep the grain
size distributions for the olivines and pyroxenes the same, as that seems the
most likely physical outcome of a collisional cascade, and introduce a
blackbody component to the fit.  The minimum size of the pyroxenes is not
well-constrained by the shape of the spectrum, because pyroxenes have
relatively broad features.  A larger minimum size for the amorphous pyroxenes
could be traded off against the blackbody grain component.

We assume that the disk dust is optically thin, so that each dust grain is in
thermal equilibrium based on its own absorption efficiency and that the various
emission components simply add.  The Spitzer 3.5, 4.5, 5.7, and 70
\micron\ photometry and the combined 5.6 -- 25 \micron\ spectrum are fit
simultaneously.

We write the emission due to the disk dust in the optically thin case
as:

\begin{equation}
F_\nu = \Sigma A_i \kappa_i B_{\nu,i}(d)
\end{equation}

where the sum is over $i$ independent dust species each with their own mass
absorption coefficients, $\kappa_i = 3 Q_{abs} /4 a \rho $ where
$Q_{abs}$ is the absorption efficiency, $a$ is the grain radius, and $\rho$ is
the grain density. The dust grains are not assumed to be in thermal contact;
thus each species has its own temperature determined by its $Q_{abs}$,
distance from the star, and the luminosity of the star by assuming it is in
thermal equilibrium.  These temperatures are calculated self-consistently by
making distance from the star the free parameter. The other free parameters are the number
or surface density of dust grains of each type, i.e. the $A_i$.

A major choice is what absorption efficiencies to use. Various authors
have taken different approaches to this problem. Because we do not leave
temperature as the free parameter, we need genuine ultraviolet and
visible absorption efficiencies, as this is where the star emits and the
grains absorb most efficiently, in order to calculate temperatures. For this
reason, we opt for Mie calculations.

Mass absorption efficiencies from the group at Kyoto Pharmaceutical
University \citep[][e.g.]{Koike:2000,Sogawa:2006} are available only in the
infrared. Optical constants from the Jena group are provided into the
UV-visible, but are only accurate in the near-infrared and longer
\citep[][e.g.]{Dorschner:1995}.  For the amorphous olivine, we adopted the
optical constants of \citet{Li:2004} which combine the Jena measurements with
ultraviolet data.  We then used a Mie code\footnotemark[1] to calculate
the mass absorption opacities.

\footnotetext[1]{http://www-atm.physics.ox.ac.uk/code/mie/index.html}

For crystalline olivine, shape is a major determinant of spectrum. The
Jena group has measured the triaxial indices of refraction, but how one
uses these to calculate the absorption is a matter of
taste. Assumption of spherical particles results in sharp peaks that are
not observed \citep{Fabian:2001}. The Jena group provides two computations of
continuous distributions of ellipsoids (CDE) in the Rayleigh limit. We
combined these with the ultraviolet efficiencies calculated in the Mie
limit from the constants of \citet{Li:2004}. We tested both the CDE1
(weighted toward spherical grains) and CDE2 (all shapes weighted
equally) calculations against our spectra and found that CDE1 was a
better match.  It is also CDE1 that fairly closely matches a real grain
measured in the Jena laboratory \citep{Fabian:2001}.

It is because of the flexibility in spectral shape granted by choosing
the exact distribution of grain shapes, that we have not attempted a
multi-component fit that would reduce our chi-square to a level
approximately equal to one. We believe that the large number of free
parameters to choose from makes this exercise of limited value.

To find the best fit, we compare the model excess as described above to the
photometric (2.16 -- 70 $\mu$m) points and IRS spectrum in the classic
$\chi^2$ formalism.  We first perform a search over a large range of parameter
space for the free parameters using the Amoeba routine from Numerical Recipes
\citep{numericalrecipes} as implemented in IDL. Once the best fit has been
found, we then loop over a small range of values for the free parameters to
determine the shape of the $\chi^2$ and determine a better minimum.

\begin{figure}
\center \includegraphics[scale=0.7]{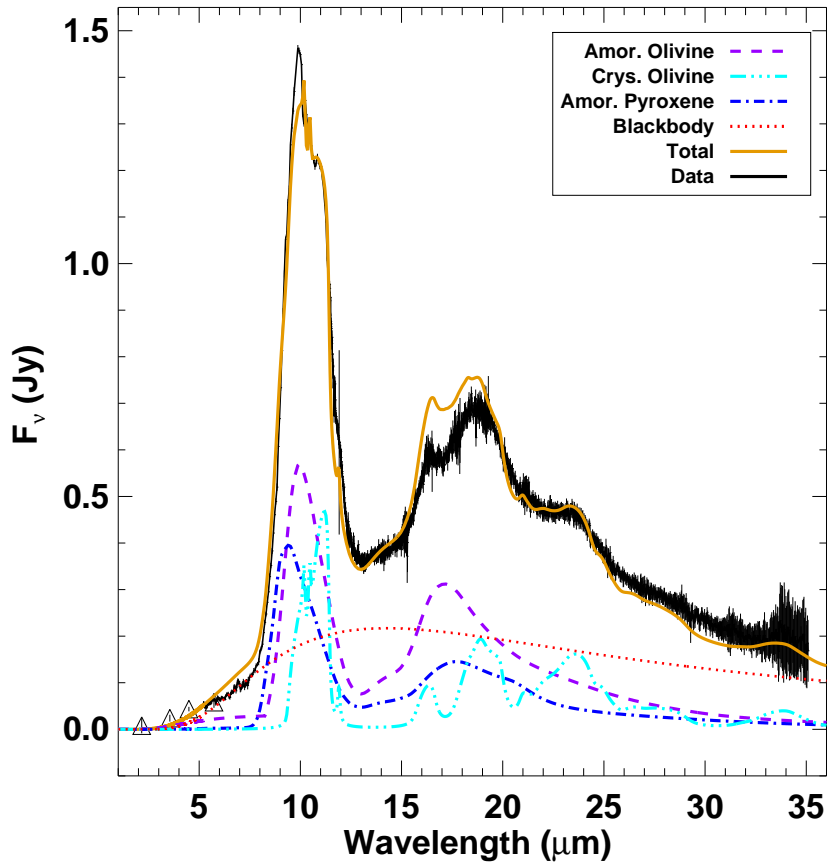}
\figcaption{Best fit model to the Spitzer spectrum (black line), derived as described in the
  text. The various dust components (purple: amorphous olivine, cyan:
  crystalline olivine, dark blue: amorphous pyroxene, and red: blackbody
  grains) sum to make the total spectrum (gold line).  All components are
  situated at the best fit distance of 0.85 AU. \label{fig_fitspectrum}}
\end{figure}

The best fit is shown in Figure \ref{fig_fitspectrum}. The best distance to
the dust is 0.85 AU from the star. The calculated temperatures of the grains
are 461 K for the pyroxene, and 460 and 507 K for the amorphous and
crystalline olivine silicates, respectively. The blackbody temperature is 358
K.  The reduced chi-square value of this fit is 19.7. Obviously, the reduced
chi-square is not near one, but nonetheless, the fit is capable of explaining
all of the main features of the spectrum including the excess emission at 3--5
\micron\, the whole of the 70 \micron\ emission, and the locations of peaks
((10, 12, 17, 19, and 24 $\mu$m) in the 7--35 \micron\ spectrum.  The overall
match between spectrum and the model is good except in the heights of the
peaks at 10 and 17 $\mu$m and the continuum at 25-30 $\mu$m. For the issues of
most relevance, namely the temperature of the dust, the presence of small
grains, and the crystallinity fraction, this model adequately represents the
spectrum.  Its predictions in the various filters are given in Table
\ref{table_photometry}. The best fit parameters are given in Table
\ref{table_model}.

The contributions by mass are 50\% amorphous olivines, 30\% amorphous
pyroxenes, and 20\% crystalline olivines. The actual masses in these small
grains range from 0.7 to 1.7 $\times$ 10$^{22}$ g. Obviously, most of the mass
would be in the blackbody grains. A minimum mass for these can be set by
setting the minimum grain radius for one of the above grains to look like a
blackbody, or $\sim$30 $\mu$m.  Then, the minimum mass is 6 $\times$ 10$^{23}$
g.

\begin{deluxetable}{llll}
\tablewidth{0pt}
  \tablecaption{Best Fit Model \label{table_model}}
  \tablehead{
  colhead{Component} &\colhead{A$_i$} &\colhead{T (K)} &\colhead{L$_{IR}$/L$_*$}}
  \startdata
  Amor. Olivine  &1.0\tablenotemark{a}     &460         &0.010\\
  Amor. Pyroxene &0.653\tablenotemark{a}   &462         &0.006\\
  Crys. Olivine  &0.821\tablenotemark{a}   &508         &0.004\\
  Blackbody      &1.64e-4                   &358         &0.011\\
  distance       &0.85 AU                  &--          & --\\
  \enddata
\tablenotetext{a}{These coefficients, defined in Section \ref{section_fitting},
  are normalized such that Amor. olivine is 1.0.}
\end{deluxetable}

\section{Discussion}

The remarkable results of the Spitzer  data are the lack of cold
dust -- all of the 70 $\mu$m emission arises from the same close-in population
of dust producing the bright 8 -- 30 $\mu$m silicate features, and that
the dust that does orbit the star is extremely close-in -- within 1 AU.  It is
the most dusty star for its age known and the one with the warmest dust.

There are now about ten known ``warm'' debris disks around FGK stars
\citep{Rhee:2008,Moor:2009,Fujiwara:2010,Melis:2010}.  Most of these have dust temperatures of
$\sim$150--200 K and L$_{IR}$/L$_*$$<$5$\times$10$^{-4}$.  For stars older
than 1 Gyr, BD +20 307 has more than two orders of magnitude more dust opacity
than the others.  The closest star in terms of dustiness is the much younger Pleiades member
HD 23514.

The timescale over which the BD+20 307 disk should evolve are short.  The
treatment of radiation pressure on small silicates in \citet{Burns:1979} and
scaled to the slightly larger mass and luminosity of BD+20 307 suggests that
all grains smaller than 0.5 \micron\ will be blown away on an orbital
timescale, which at 0.85 AU, assuming a central mass of 2.2 M$_\odot$, is 0.5
yr. 

Taking the model fit L$_{IR}$/L$_*$ = 0.032 as the surface density of the
grains yields a collision time of 2.4 yr. This is only a few times longer than
the orbital time, so the small grains observed in the spectrum will be created and then lost nearly
immediately.  \citet{Wyatt:2007} argue that massive planetsimal belts grind
themselves down by collisions quickly, so it is not possible to maintain very
large amounts of dust in the terrestial planet region of old stars. They
included BD+20 307 in the list of stars for which it was impossible to
generate the present dust through collisional grinding, even using the IRAS 60
$\mu$m upper limit to constrain the cold planetesimal population and for an age of 300
Myr. At the current best estimate of its age of at least 1000 Myr, its maximum
\lir\ in steady state (see their Eq. 20) is 7.4 $\times$ 10$^{-8}$.  Recently,
\cite{Heng:2010} confirm the \citet{Wyatt:2007} result that BD +20 307 cannot
be from a steady-state collisional cascade, although their models allow most
of the other warm disk sources to come from a warm planetesimal cascade.

\begin{figure}
\center \includegraphics[angle=90,scale=0.7]{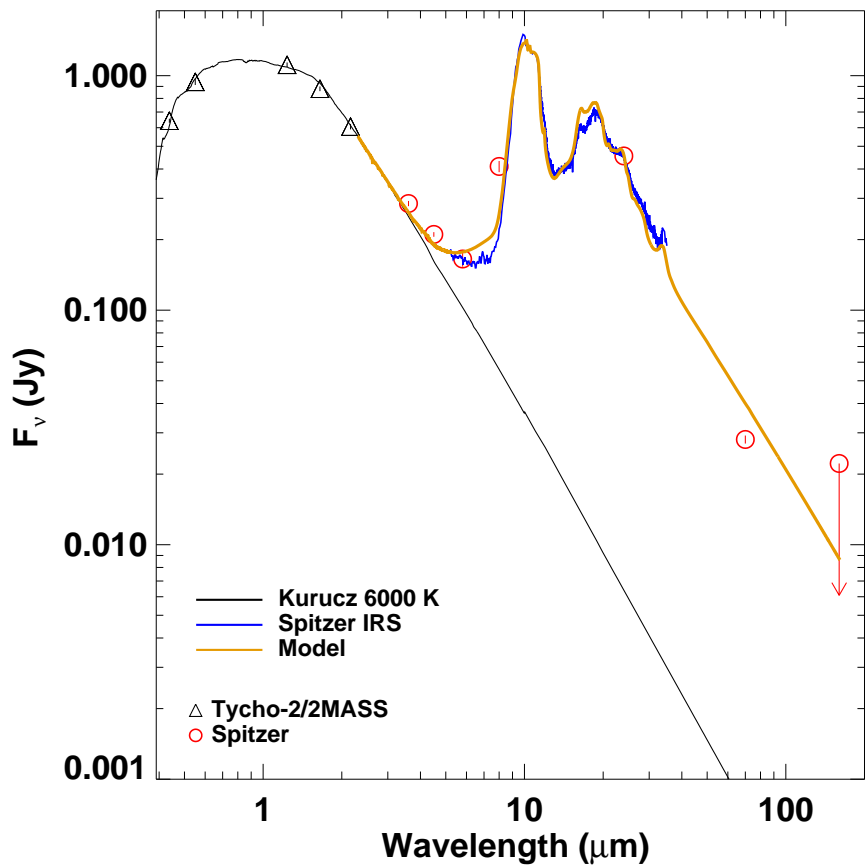}
\figcaption{Spectral energy distribution of BD+20 307 showing the visual and
  near-infrared photometry from the Tycho-2 and 2MASS catalogs along with the
  Spitzer photometry and spectroscopy from Figure 1 and our best-fit model
  from Figure 2. We do not detect any cold dust; our best-fit model for the hot dust actually
overpredicts the measured 70 $\mu$m flux density and is slightly higher than
the 1 $\sigma$ upper limit at 160 $\mu$m (the 3 $\sigma$ upper limit is
plotted).  The model does a reasonable job of reproducing the data for the
issues we care about, namely, the dust location (temperature), the presence of small grains, and
the crystallinity fraction.\label{fig_sedwithmodel}}
\end{figure}

We do not detect any cold dust; our best-fit model for the hot dust actually
overpredicts the measured 70 $\mu$m flux density and is slightly higher than
the 1 $\sigma$ upper limit at 160 $\mu$m (Figure \ref{fig_sedwithmodel}).  A
check on the reasonableness of the lack of cold dust, is that if we take the
average flux density in the IRS spectrum of 0.4 Jy and extrapolate this as a
Rayleigh-Jeans fall-off from the mean wavelength of the IRS spectrum, 20
$\mu$m, we would predict a 70 $mu$m flux density of 0.032 Jy, which is again
higher than measured.

We can place a conservative upper limit on the amount of cold dust by finding the
luminosity of a dust ring whose spectrum peaks at 160 $\mu$m (i.e. temperature
of 33 K) and produces no more than a 10\% increase in the flux density at 70
$\mu$m -- such a cold disk would produce an additional 6~mJy at 160 $\mu$m,
which would still fall under the measured 3 $\sigma$ upper limit and would
have a L$_{IR}$/L$_*$ = 3 $\times$ 10$^{-5}$.  Therefore, in the context of
the Wyatt et al. model for planetesimal belt self-grinding, the lack of
detectable cold dust rules out any belt within 200 AU that is losing mass at a
rate sufficient to account for the hot dust.

Late heavy bombardment models per se are not attractive analogies for BD+20
307. In the \citep{Gomes:2005} scenario, a late crossing of the 2:1 resonance
between Jupiter and Saturn swept resonances across the asteroid and Kuiper
belts and dislodged many small bodies that then impacted the terrestrial
planets. For a similar scenario to work for BD+20 307, the resonance crossing
must be delayed even longer, until 1 Gyr, but that is likely achievable by
tuning the initial configuration of planets.  More problematic is that the Kuiper
Belt prior to the late heavy bombardment, which was massive enough to generate
giant planet migration, would itself produce significant cold dust
\citep{Meyer:2007}. 

The remaining hypothesis for the source of the hot dust around BD+20 307 is a
giant impact.  Detailed models exist of one such impact -- the Moon-forming
event.  About a quarter of the material that forms the circum-terrestrial disk
out of which the Moon formed starts very hot, at temperatures $>$1000 K, or
high enough to vaporize silicates and metals \citep{Canup:2004}. Furthermore,
a few percent or more of a Lunar mass can escape the Earth-Moon system.  The
small particles around BD+20 307 are $\sim$20\% crystalline. This level of
crystallinity is well within range of T Tauri star dust; in contrast,
differentiated bodies such as large asteroids or planets should be nearly
100\% crystalline.  That the observed dust is largely amorphous suggests that
the impact heated the bulk of the material to temperatures $\sim$1000~K and
that any recondensation took place under low vapor pressure conditions where
the molten silicate cooled quickly and therefore did not re-anneal into
crystalline form.

A terrestrial planet system may be stable over a Gyr but not permanent.  The
Planet V models of \citet{Chambers:2007} show that in a configuration such as
the Solar System's, a fifth planet can survive in the inner Solar System for a
Gyr and create collisions of terrestrial planets at many hundreds of millions
of years after the system formed. The configuration of the exoplanet system
will determine how long it lasts; small planets further from the star take
longer to go unstable (Chambers, personal communication).  In the case where a
tenuous asteroid belt still remains in a planetary system, the trajectory of a
rogue planet can destabilize the belt and produce collisions between
planetesimals or bombardment of the planets.  Over several Gyr, our current
terrestrial planet system may be unstable and result in collisions between the
planets \citep{Laskar:2009}.

Recently, \citet{Matranga:2010} suggested that the changing separation of close
binaries due to angular momentum loss could sweep resonances across
planetary systems and destabilize their planetesimal belts.  The stars in
BD+20 307 are separated by 0.06 AU at present \cite{Weinberger:2008}, and thus
the disk radius is 14$\times$ the binary separation. They have likely been in
a tight circular orbit for billions of years \cite{Weinberger:2008} and their
relatively low X-ray activity likely makes their angular momentum loss
relatively slow. 

How common are giant collisions amongst sun-like stars? We can estimate the
number of giant impacts per star $N_g = f_* A L^{-1}$ where $f_*$ is the
fraction of stars observed to have hot dust, $L$ is the lifetime of the
collision products, and $A$ is the age of the stars surveyed.

To estimate the lifetime of the dust, we invert the \citep{Wyatt:2007}
expression for the amount of dust that can be produced by a collisional
cascade over time (their equation 20) to solve for the lifetime of a collision
that starts in a giant impact. The most significant unknown is the internal
strength of the large planetesimal(s) involved, but the binding energy is
likely closer to that of a differentiated planet than a pure rubble pile,
i.e. more like 2$\times$10$^5$ than 200 J kg$^{-1}$.  Therefore, a
planet-sized impact could generate a cascade that lasts up to 80,000 yr. This
result is similar to the 100,000 yr lifetime for a collisional cascade started
by planetary embryos as calculated by \citep{Melis:2010}.

Surveys with Spitzer have looked at all Sun-like stars within 25 pc
\cite{Koerner:2010}, for a sample of about 800. The average age of this
population is $\gtrsim$3 Gyr but it includes a small fraction of much younger
stars \citep{Wright:2004}.  While there are five stars with warm dust in this
sample, we note that these have 24 $\mu$m flux densities $\lesssim$10 times
their photospheres and \lir\ $\lesssim$ 10$^{-3}$, and so are not nearly as
extreme as BD+20 307. Moreover, their average age is $<$1 Gyr.  Their warm
dust may either be from young planetesimal belts (e.g. $\beta$ Pic, which is
in this group) or from collisions fed by outer belts and constrained by
planets (e.g. $\eta$ Corvi; Wyatt et al. 2007).

IRAS was capable of detecting every A--G star with as much hot dust as BD+20
307 out to $\sim$280 pc, although it is possible some of these would not have
been noticed. There are $\sim$600,000 stars stars of earlier spectral type
than K0 in that volume.  Four stars (including BD +20 307) have F$_{24}$/F$_*$
$>$ 10. This fraction of host dust stars implies a giant impact rate greater
than 0.2 impacts/star during its main sequence lifetime, i.e. implies that
such impacts are common.  Obviously, these are small number statistics. The
WISE mission currently surveying the sky at mid-infrared wavelengths will
greatly improve the statistics. Its sensitivity is 500 times better than IRAS
at $\sim$24 $\mu$m \citep{Liu:2008}, and so it should provide a truly complete census of
isolated hot dust sources out to a few kpc.

It is difficult to search for planets around BD+20 307 with the radial
velocity technique because of the spectroscopic binary.  However, if planets
do encircle this star, they may trap small bodies in resonance.  An
enhancement in the planetesimal density might increase the liklihood of a
destructive collision such as the one observed.  The star HD 69830, also an
IRAS source, and $>$ 1 Gyr old, was shown with Spitzer to harbor a ring very
similar in size and temperature to that around BD+20 307, but containing 150
times less dust \citep{Beichman:2005}. A system of three Neptune mass planets
was then found to encircle HD 69830, with the dust either residing between the
2nd and 3rd planets from the star or exterior to all three \citep{Lovis:2006}.

Whatever the scenario that orginated the collision in the BD+20 307 system and
others, it is interesting to note that the spectra of other warm dust systems
are not identical. HD 69830, for example, has much more prominent 20 $\mu$m
crsytalline silicate peaks \citep{Beichman:2005} and HD 23514 has an unusal
broad spectral feature at 9 rather than 10 $\mu$m \citep{Rhee:2008}. HD
172555's narrow 8 $\mu$m peak and featureless 20$\mu$m regions have been
attributed to SiO.  These various compositions may result from a combination
of progenitor composition, which would affect the primordial abundances of
different types of silicates, and geometry of the impact, which can change the
melt fraction and condensation properties of its debris.

\section{Conclusions}

We have detected a large amount of small silicate dust at 0.85 AU from the
$>$Gyr old star BD+20 307 but no cold dust that could be generated tens or
hundreds of AU from the star. A catastrophic collision of two rocky,
planetary-scale bodies in the terrestrial zone is the most likely source for this
warm dust because it does not require a reservoir of planetesimals in the
outer system. Furthermore, the high amorphous silicate content of the dust
suggests that the impact must have caused a high melt fraction and
amorphization of what was likely a largely crystalline parent body.

\acknowledgements

This work is based on observations made with the Spitzer Space Telescope,
which is operated by the Jet Propulsion Laboratory, California Institute of
Technology under a contract with NASA. Support for this work was provided by
NASA through an award issued by JPL/Caltech and a contract to Ames Research
Center for the SOFIA program. We acknowledge support from NASA's Astrobiology
Institute to the UCLA and CIW nodes. John Chambers and Sarah Stewart provided
insightful conversations on the topic of giant impacts.

\bibliographystyle{apj}

\begin{thebibliography}{36}
\expandafter\ifx\csname natexlab\endcsname\relax\def\natexlab#1{#1}\fi

\bibitem[{Ancker {et~al.}(2001)Ancker, Meeus, Cami, Waters, \&
  Waelkens}]{Ancker:2001}
Ancker, M. E. V.~D., Meeus, G., Cami, J., Waters, L. B. F.~M., \& Waelkens, C.
  2001, A{\&}A, 369, L17

\bibitem[{Beichman {et~al.}(2005)Beichman, Bryden, Gautier, Stapelfeldt,
  Werner, Misselt, Rieke, Stansberry, \& Trilling}]{Beichman:2005}
Beichman, C.~A., Bryden, G., Gautier, T.~N., Stapelfeldt, K.~R., Werner, M.~W.,
  Misselt, K., Rieke, G., Stansberry, J., \& Trilling, D. 2005, ApJ, 626, 1061

\bibitem[{Benz {et~al.}(1988)Benz, Slattery, \& Cameron}]{Benz:1988}
Benz, W., Slattery, W.~L., \& Cameron, A. G.~W. 1988, Icarus, 74, 516

\bibitem[{Burns {et~al.}(1979)Burns, Lamy, \& Soter}]{Burns:1979}
Burns, J.~A., Lamy, P.~L., \& Soter, S. 1979, Icarus, 40, 1

\bibitem[Campins \& Ryan(1989)]{Campins1989} Campins, H., \& Ryan, E.~V.\ 1989, \apj, 341, 1059 

\bibitem[{Canup(2004)}]{Canup:2004}
Canup, R.~M. 2004, Icarus, 168, 433

\bibitem[{Chambers(2007)}]{Chambers:2007}
Chambers, J.~E. 2007, Icarus, 189, 386

\bibitem[{Dorschner {et~al.}(1995)Dorschner, Begemann, Henning, Jaeger, \&
  Mutschke}]{Dorschner:1995}
Dorschner, J., Begemann, B., Henning, T., Jaeger, C., \& Mutschke, H. 1995,
  A{\&}A, 300, 503

\bibitem[{Fabian {et~al.}(2001)Fabian, Henning, J�Ger, Mutschke, Dorschner,
  \& Wehrhan}]{Fabian:2001}
Fabian, D., Henning, T., J�Ger, C., Mutschke, H., Dorschner, J., \& Wehrhan,
  O. 2001, A{\&}A, 378, 228

\bibitem[{Fujiwara {et~al.}(2010)Fujiwara, Onaka, Ishihara, Yamashita,
  Fukagawa, Nakagawa, Kataza, Ootsubo, \& Murakami}]{Fujiwara:2010}
Fujiwara, H., Onaka, T., Ishihara, D., Yamashita, T., Fukagawa, M., Nakagawa,
  T., Kataza, H., Ootsubo, T., \& Murakami, H. 2010, ApJ, 714, L152

\bibitem[{Gomes {et~al.}(2005)Gomes, Levison, Tsiganis, \&
  Morbidelli}]{Gomes:2005}
Gomes, R., Levison, H.~F., Tsiganis, K., \& Morbidelli, A. 2005, Nature, 435,
  466

\bibitem[{Gordon {et~al.}(2007)Gordon, Engelbracht, Fadda, Stansberry, Wachter,
  Frayer, Rieke, Noriega-Crespo, Latter, Young, Neugebauer, Balog, Beeman,
  Dole, Egami, Haller, Hines, Kelly, Marleau, Misselt, Morrison,
  P{\'e}rez-Gonz{\'a}lez, Rho, \& Wheaton}]{Gordon:2007}
Gordon, K.~D., Engelbracht, C.~W., Fadda, D., Stansberry, J., Wachter, S.,
  Frayer, D.~T., Rieke, G., Noriega-Crespo, A., Latter, W.~B., Young, E.,
  Neugebauer, G., Balog, Z., Beeman, J.~W., Dole, H., Egami, E., Haller, E.~E.,
  Hines, D., Kelly, D., Marleau, F., Misselt, K., Morrison, J.,
  P{\'e}rez-Gonz{\'a}lez, P., Rho, J., \& Wheaton, W.~A. 2007, PASP, 119, 1019

\bibitem[{Greaves {et~al.}(2009)Greaves, Stauffer, Cameron, Meyer, \&
  Sheehan}]{Greaves:2009}
Greaves, J.~S., Stauffer, J.~R., Cameron, A.~C., Meyer, M.~R., \& Sheehan, C.
  K.~W. 2009, MNRAS, 394, L36

\bibitem[{Heng \& Tremaine(2010)}]{Heng:2010}
Heng, K. \& Tremaine, S. 2010, MNRAS, 401, 867

\bibitem[{Kenyon \& Bromley(2004)}]{Kenyon:2004}
Kenyon, S.~J. \& Bromley, B.~C. 2004, ApJ, 602, L133

\bibitem[{Koerner {et~al.}(2010)Koerner, Kim, Trilling, Larson, Cotera,
  Stapelfeldt, Wahhaj, Fajardo-Acosta, Padgett, \& Backman}]{Koerner:2010}
Koerner, D.~W., Kim, S., Trilling, D.~E., Larson, H., Cotera, A., Stapelfeldt,
  K.~R., Wahhaj, Z., Fajardo-Acosta, S., Padgett, D., \& Backman, D. 2010, ApJ,
  710, L26

\bibitem[{Koike {et~al.}(2000)Koike, Tsuchiyama, Shibai, Suto, Tanab{\'e},
  Chihara, Sogawa, Mouri, \& Okada}]{Koike:2000}
Koike, C., Tsuchiyama, A., Shibai, H., Suto, H., Tanab{\'e}, T., Chihara, H.,
  Sogawa, H., Mouri, H., \& Okada, K. 2000, A{\&}A, 363, 1115

\bibitem[{Laskar \& Gastineau(2009)}]{Laskar:2009}
Laskar, J. \& Gastineau, M. 2009, Nature, 459, 817

\bibitem[{Li {et~al.}(2004)Li, Zhao, \& Li}]{Li:2004}
Li, M., Zhao, G., \& Li, A. 2004, ApJ, 613, L145

\bibitem[{Lisse {et~al.}(2009)Lisse, Chen, Wyatt, Morlok, Song, Bryden, \&
  Sheehan}]{Lisse:2009}
Lisse, C.~M., Chen, C.~H., Wyatt, M.~C., Morlok, A., Song, I., Bryden, G., \&
  Sheehan, P. 2009, ApJ, 701, 2019

\bibitem[{Liu {et~al.}(2008)}]{Liu:2008}Liu, F. et al. 2008, \procspie, 7017

\bibitem[{Lovis {et~al.}(2006)Lovis, Mayor, Pepe, Alibert, Benz, Bouchy,
  Correia, Laskar, Mordasini, Queloz, Santos, Udry, Bertaux, \&
  Sivan}]{Lovis:2006}
Lovis, C., Mayor, M., Pepe, F., Alibert, Y., Benz, W., Bouchy, F., Correia, A.
  C.~M., Laskar, J., Mordasini, C., Queloz, D., Santos, N.~C., Udry, S.,
  Bertaux, J.-L., \& Sivan, J.-P. 2006, Nature, 441, 305

\bibitem[{Marinova {et~al.}(2008)Marinova, Aharonson, \&
  Asphaug}]{Marinova:2008}
Marinova, M.~M., Aharonson, O., \& Asphaug, E. 2008, Nature, 453, 1216

\bibitem[Matranga et al.(2010)]{Matranga:2010} Matranga, M., Drake, 
J.~J., Kashyap, V.~L., Marengo, M., 
\& Kuchner, M.~J.\ 2010, \apjl, 720, L164 

\bibitem[Melis et al.(2010)]{Melis:2010} Melis, C., Zuckerman, B., 
Rhee, J.~H., \& Song, I.\ 2010, \apjl, 717, L57 



\bibitem[{Meyer {et~al.}(2007)Meyer, Backman, Weinberger, \&
  Wyatt}]{Meyer:2007}
Meyer, M.~R., Backman, D.~E., Weinberger, A.~J., \& Wyatt, M.~C. 2007,
  Protostars and Planets V, 573

\bibitem[{Meyer {et~al.}(2008)Meyer, Carpenter, Mamajek, Hillenbrand,
  Hollenbach, Moro-Martin, Kim, Silverstone, Najita, Hines, Pascucci, Stauffer,
  Bouwman, \& Backman}]{Meyer:2008}
Meyer, M.~R., Carpenter, J.~M., Mamajek, E.~E., Hillenbrand, L.~A., Hollenbach,
  D., Moro-Martin, A., Kim, J.~S., Silverstone, M.~D., Najita, J., Hines,
  D.~C., Pascucci, I., Stauffer, J.~R., Bouwman, J., \& Backman, D.~E. 2008,
  ApJ, 673, L181

\bibitem[{Mo{\'o}r {et~al.}(2009)Mo{\'o}r, Apai, Pascucci, {\'A}brah{\'a}m,
  Grady, Henning, Juh{\'a}sz, Kiss, \& K{\'o}sp{\'a}l}]{Moor:2009}
Mo{\'o}r, A., Apai, D., Pascucci, I., {\'A}brah{\'a}m, P., Grady, C., Henning,
  T., Juh{\'a}sz, A., Kiss, C., \& K{\'o}sp{\'a}l, {\'A}. 2009, ApJ, 700, L25

\bibitem[{{Press} {et~al.}(1992){Press}, {Teukolsky}, {Vetterling}, \&
  {Flannery}}]{numericalrecipes}
{Press}, W.~H., {Teukolsky}, S.~A., {Vetterling}, W.~T., \& {Flannery}, B.~P.
  1992, {Numerical recipes in C. The art of scientific computing}, 2nd edn.
  (Cambridge: University Press)

\bibitem[{Rhee {et~al.}(2008)Rhee, Song, \& Zuckerman}]{Rhee:2008}
Rhee, J.~H., Song, I., \& Zuckerman, B. 2008, ApJ, 675, 777

\bibitem[{Sogawa {et~al.}(2006)Sogawa, Koike, Chihara, Suto, Tachibana,
  Tsuchiyama, \& Kozasa}]{Sogawa:2006}
Sogawa, H., Koike, C., Chihara, H., Suto, H., Tachibana, S., Tsuchiyama, A., \&
  Kozasa, T. 2006, A{\&}A, 451, 357

\bibitem[{Song {et~al.}(2005)Song, Zuckerman, Weinberger, \&
  Becklin}]{Song:2005}
Song, I., Zuckerman, B., Weinberger, A.~J., \& Becklin, E.~E. 2005, Nature,
  436, 363

\bibitem[{Strom {et~al.}(2005)Strom, Malhotra, Ito, Yoshida, \&
  Kring}]{Strom:2005}
Strom, R.~G., Malhotra, R., Ito, T., Yoshida, F., \& Kring, D.~A. 2005,
  Science, 309, 1847

\bibitem[{Su {et~al.}(2006)Su, Rieke, Stansberry, Bryden, Stapelfeldt,
  Trilling, Muzerolle, Beichman, Moro-Martin, Hines, \& Werner}]{Su:2006}
Su, K. Y.~L., Rieke, G.~H., Stansberry, J.~A., Bryden, G., Stapelfeldt, K.~R.,
  Trilling, D.~E., Muzerolle, J., Beichman, C.~A., Moro-Martin, A., Hines,
  D.~C., \& Werner, M.~W. 2006, ApJ, 653, 675

\bibitem[{Trilling {et~al.}(2008)Trilling, Bryden, Beichman, Rieke, Su,
  Stansberry, Blaylock, Stapelfeldt, Beeman, \& Haller}]{Trilling:2008}
Trilling, D.~E., Bryden, G., Beichman, C.~A., Rieke, G.~H., Su, K. Y.~L.,
  Stansberry, J.~A., Blaylock, M., Stapelfeldt, K.~R., Beeman, J.~W., \&
  Haller, E.~E. 2008, ApJ, 674, 1086

\bibitem[{van Leeuwen(2007)}]{Leeuwen:2007}
van Leeuwen, F. 2007, A{\&}A, 474, 653

\bibitem[{Weinberger(2008)}]{Weinberger:2008}
Weinberger, A.~J. 2008, ApJ, 679, L41

\bibitem[{Wright {et~al.}(2004)Wright, Marcy, Butler, \& Vogt}]{Wright:2004}
Wright, J.~T., Marcy, G.~W., Butler, R.~P., \& Vogt, S.~S. 2004, ApJS, 152, 261

\bibitem[{Wyatt {et~al.}(2007)Wyatt, Smith, Greaves, Beichman, Bryden, \&
  Lisse}]{Wyatt:2007}
Wyatt, M.~C., Smith, R., Greaves, J.~S., Beichman, C.~A., Bryden, G., \& Lisse,
  C.~M. 2007, ApJ, 658, 569

\bibitem[{Zuckerman {et~al.}(2008)Zuckerman, Fekel, Williamson, Henry, \&
  Muno}]{Zuckerman:2008}
Zuckerman, B., Fekel, F.~C., Williamson, M.~H., Henry, G.~W., \& Muno, M.~P.
  2008, ApJ, 688, 1345

\end{thebibliography}

\end{document}